\documentclass{PoS}
\usepackage{gensymb}
\usepackage{subcaption}
\usepackage{amsmath}
\title{Testing the Limits of Particle Acceleration in Cygnus OB2 with HAWC}

\ShortTitle{Cygnus Cocoon}

\author{\speaker{B. Hona}\\
        Michigan Technological University\\
        E-mail: \email{bhona@mtu.edu}}
        
\author{{H. Fleischhack}\\
        Michigan Technological University\\
        }
        
\author{{ P. Huentemeyer}\\
        Michigan Technological University\\
        }
        
\author{{for the HAWC Collaboration}\footnote{for collaboration list see PoS(ICRC2019)1177}\\
For a complete author list and acknowledgement see { \href{https://www.hawc-observatory.org/collaboration/icrc2019.php}{https://www.hawc-observatory.org/collaboration/icrc2019.php}}
}

\abstract{Star forming regions (SFRs) have been postulated as possible sources of cosmic rays (CRs) in our galaxy. One example of a gamma-ray source associated with an SFR is the Fermi-LAT cocoon, an extended region of gamma-ray emission in the Cygnus X region and attributed to a possible superbubble with freshly accelerated CRs. Because the emission region is surrounded by ionization fronts, it has been named the $Cygnus\ cocoon$. CRs in the cocoon could have originated in the OB2 association and been accelerated at the interaction sites of stellar winds of massive O type stars. So far, there is no clear association at TeV energies. Spectral and morphological studies of TeV gamma-ray emission detected by the High Altitude Water Cherenkov (HAWC) observatory at the 2HWC J2031+415 region reveal that the spectral energy distribution of the cocoon extends from GeV to at least tens of TeV. Using HAWC data, we are able to study the acceleration of particles to highest energies in the Cygnus OB2 SFR.}
          
\FullConference{36th International Cosmic Ray Conference -ICRC2019-\\
		July 24th - August 1st, 2019\\
		Madison, WI, U.S.A.}

\begin{document}

\section{Introduction}

Cosmic rays in our Galaxy can reach up to a few PeV in energy. However, the Galactic astrophysical accelerators which are associated with majority of the highest energy gamma rays have not been understood. Apart from the proposition of a supermassive black hole as a CR accelerator for PeV particles at the Galactic centre by the HESS observatory \cite{pevatron}, no other source or source type has been identified as a PeV accelerator. \\ 

Supernova Remnants (SNRs) have been postulated as the main sources of CRs in our Galaxy and CRs can be accelerated to hundress of TeV in SNR environments \cite{butt}. The cutoffs in the gamma-ray spectra of SNRs at several TeV show that until now there is no evidence that SNRs accelerate particles beyond a few hundreds of TeV \cite{julia}.  Along with SNRs, star forming regions (SFRs) have been  speculated as  possible CR acceleration sites \cite{sfr}.  As the SFR evolves into stellar clusters and associations, CRs can be accelerated via two main ways. The interaction of supersonic winds of the massive stars in the clusters creates a collective bubble called a superbubble. Strong shock waves are formed at the interaction sites of the stellar winds \cite{bykov}. Secondly, Supernova (SN) explosions of massive stars (M$\geq$ 8\(M_\odot\)) after a few Myrs can contribute kinetic energy to accelerate charged particles \cite{hess}. Collective stellar winds and SN explosions in the stellar clusters can accelerate electrons and protons up to TeV energies \cite{SFRromero}. \\

The Cygnus region (70\degree{}< l < 85\degree{} and -4\degree{}< b < 4\degree) consists of a number of stellar clusters and associations. A particularly interesting site is OB2 association, one of the most massive stellar associations in our Galaxy with $\sim$ 120 type O stars \cite{OB2mass}. It is possibly responsible for the extended region of hard GeV emission seen at this location by Fermi-LAT observatory, known as the Cygnus cocoon  \cite{Acker11}.The Cocoon, detected in the energy range of (1--100) GeV has no counterpart at lower wavelengths. Observations in the (2--10) keV by Suzaku observatory concluded that extended emission detected at the location after subtraction of known sources from X-ray images and Cosmic X-ray background is related to Galactic ridge X-ray emission rather than the Cocoon \cite{suzaku}. In the TeV range, the emission seen by MILAGRO, ARGO and the HAWC observatory might be related to GeV Cocoon \cite{abdo12, Bartoli14, catalog}. Using the observation at higher TeV energies by HAWC, we can learn about the physics processes that accelerate the CRs to hundreds of TeV and also constrain the lower limit to the maximum CR accerelation by OB2 association. This analysis focuses on spectral and morphological studies of the cocoon region to disentangle TeV gamma-ray emission detected by the HAWC and the other gamma-ray instruments and to understand the origin of the emission.

\section{Fermi-LAT Cocoon and OB2 Association}
Fermi-LAT is a satellite-based gamma-ray observatory, sensitive to gamma-ray emission in the 20 MeV to beyond 500 GeV range\footnote{\texttt{https://fermi.gsfc.nasa.gov/science/instruments/table1-1.html}}. The observatory detected 50 pc wide extended excess of gamma-ray emission in the Cygnus region after subtraction of the interstellar background and known sources in the region  \cite{Acker11}. This excess emission is detected with $10.1\sigma $ significance above 1 GeV and is attributed to energetic particles inside the interstellar cavities created as a result of stellar activities \cite{Acker11}. The hard spectrum ($\sim$ -2.1) observed for this GeV emission indicates the source of the emission to be freshly-accelerated cosmic rays  \cite{Acker11}. \\

The cocoon lies in the bright Cygnus X region between the OB2 association and the SNR gamma Cygni. Either or both of these objects could be the sources of the CRs in the cocoon. The Fermi-LAT study explored the possibility of the 7000 year old gamma Cygni SNR as a possible source of the cocoon. However, the morphology of this SNR and the lack of evidence of a shockwave from gamma Cygni toward the direction of the cocoon seem to rule out gamma Cygni as a possible accelerator for the cocoon \cite{Acker11}. Another possible source for CRs in the cocoon is the OB2 association. The age of the stars in the association ranges from $3.5_{-1.0}^{+0.8}$ Myr in the core of the association to $5.3_{-1.0}^{+1.5}$ Myr in the northwest \cite{age}. Collective effects of stellar wind interactions of massive type O stars in the OB2 association could have accelerated the CRs that make up the cocoon. Thus, this provides a rare opportunity to study production and acceleration mechanisms of cosmic rays in our Galaxy within the SFR. Assuming that the accelerated particles in the cocoon are protons and originated from the OB2 association, the Fermi-LAT collaboration calculated that the energetic protons in the cocoon can remain confined for over 100,000 years. This result matches the time scale given by isotopic abundances \cite{Acker11}. Hence, the source detected by Fermi-LAT is possibly an active CR superbubble \cite{Acker11}. Protons trapped in this superbubble interact with the surrounding molecular cloud nuclei resulting in gamma-ray emission detected by the Fermi-LAT instrument.

\begin{figure}
\begin{subfigure}{.5\textwidth}
\captionsetup{width=.8\linewidth}
\includegraphics[scale=0.5]{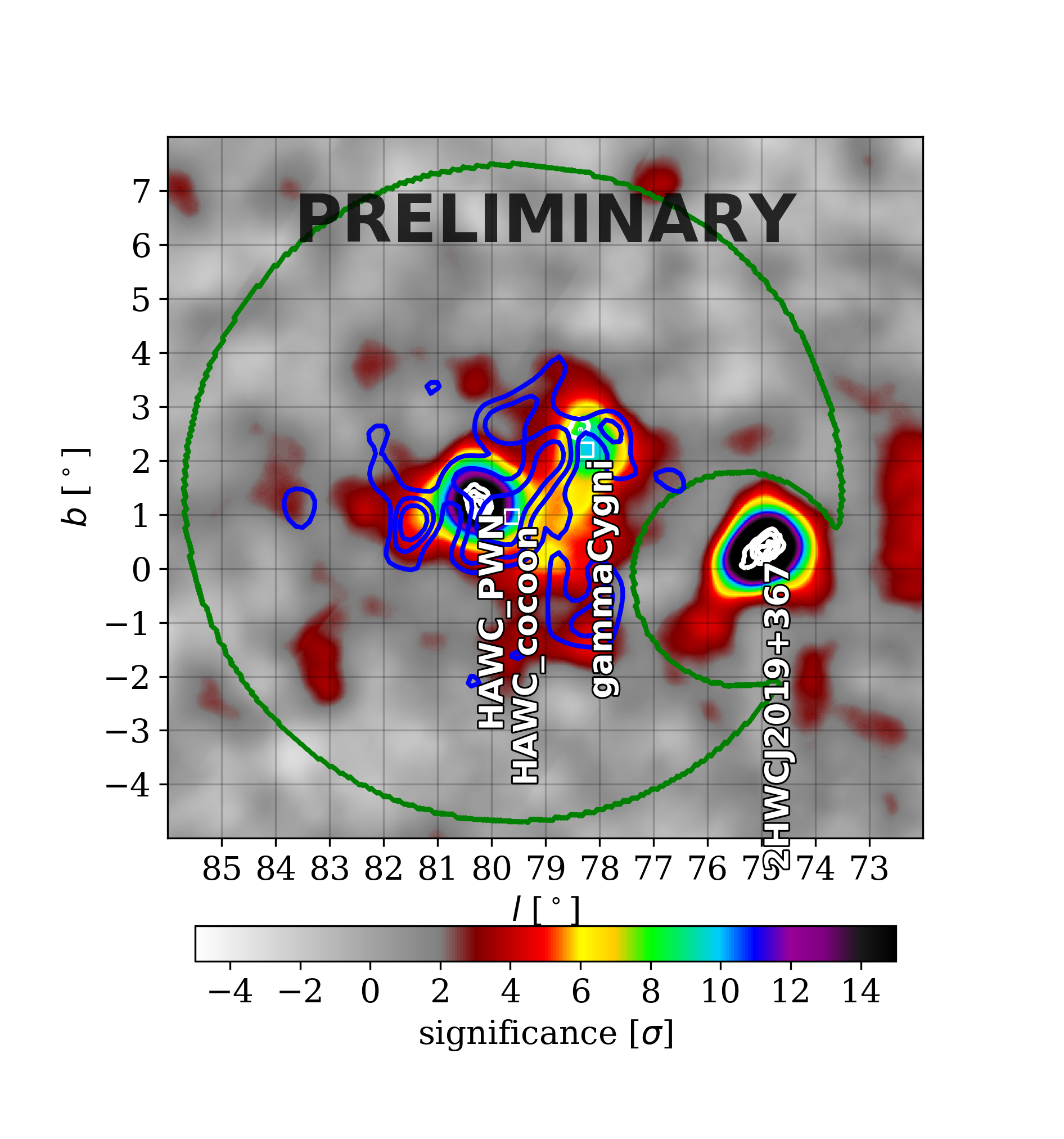}
\caption{Significance map of the cocoon region with 0.5\degree extended source assumption and 1038 days of HAWC data. The white contours are the VERITAS significance contours for 5, 7, 9 and 11 $\sigma$. Green line encloses the region of interest (ROI) for this study.}
\label{fig:data}
\end{subfigure}
\begin{subfigure}{.5\textwidth}
\captionsetup{width=.8\linewidth}
\includegraphics[scale=0.5]{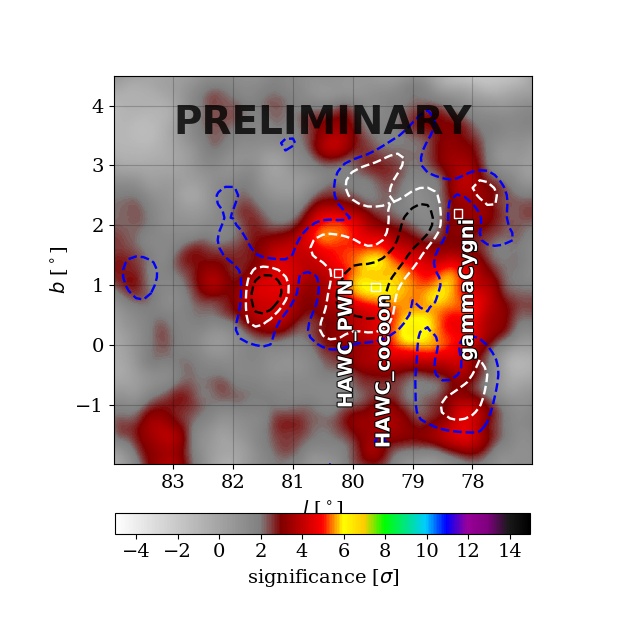}
\caption{Significance map of the excess emission after subtracting all the other sources (PWN and gammaCygni) in the region and with 0.5\degree smearing applied. The blue, white and the black contours are published Fermi-LAT observation of 0.16, 0.24, 0.32 photons/bin respectively.}
\label{fig:cocoon}
\end{subfigure}
\captionsetup{width=.9\linewidth}
\caption{Significance maps of the cocoon region. An extended emission co-located with the Fermi-LAT cocoon is significantly detected after the subtraction of other sources \& background emission.}
\end{figure}

\section{HAWC Observatory}
The HAWC observatory is a wide field TeV gamma-ray observatory located at Sierra Negra, Mexico at an altitude of 4100 m \cite{Abey13a}. The HAWC gamma-ray instrument is comprised of an array of 300 water Cherenkov detectors (WCDs) and is sensitive to gamma rays in the energy range of a few 100 GeV to beyond 100 TeV \cite{hawc2019crab}.  Each WCD has four photomultipliers tubes (PMT) at the bottom which detect the Cherenkov light produced via charged particles from the air showers travelling in the WCD. Air shower events recorded by the detector are recontructed to extract shower properties and after this reconstruction process, event and background maps are generated \cite{Abey17crab}. The hadronic cosmic rays that pass gamma/hadron separation cuts during reconstruction form the main background in the analysis of gamma-ray sources. The background for each bin is estimated using the method of "direct integration" \cite{Abey17crab}.  \\

HAWC data is divided into 9 size bins according to the fraction of PMTs triggered. Size bin 9 has the highest energy gamma rays and best angular resolution (68\% containment radius) of 0.17\degree \cite{Abey17crab}. Using the newly developed energy estimation algorithms, each size bin is further subdivided into 12 quarter decade energy bins for a total of 108 bins \cite{hawc2019crab}. Currently HAWC utilises two different methods of estimating gamma-ray energy. One is called the "ground parameter" which uses charge density at 40 m from the shower axis \cite{kelly}. Another method is called the "neural network" which relies on an artificial neural network and various parameters from the  event reconstrunction process \cite{jim}. For this anaylysis, studies were done with both methods to provide cross checks and to better understand the detector systematics.  Reconstructed energy bins higher than 1 TeV and 1128 days HAWC data were used. The results presented here are with the "ground parameter" method. These results agree with the studies from the "neural network" method within statistical uncertainties.

\section{The Cocoon Region}
Shown here in Fig.~\ref{fig:data} is the significance map of the Cocoon region with 1038 days of HAWC data using the "ground parameter" method and with 0.5\degree{} extended source assumption and a spectral index of -2.7. 2HWC J2031+415 is one of the bright Cygnus region sources listed in the second HAWC catalog which was detected with significance of 14.69 $\sigma$ at the location (RA =307.93\degree, Dec=41.51\degree){} \cite{catalog}. The catalog used 9 size bins without subdivision into energy bins and 507 days of data. The 2HWC J2031+415 region is co-located with two sources, the Fermi-LAT Cocoon at GeV energies shown in the blue contours in Fig.~\ref{fig:data}, and VER J2031+415 \cite{veritas2014}, a TeV source observed by the VERITAS observatory, shown in the white contours in Fig.~\ref{fig:data}. Using the ground parameter energy estimator, we were able to disentangle the contribution of these two overlapping sources to the TeV gamma-ray emission observed at  the 2HWC J2031+415 region. The ROI used (green contour in Fig.~\ref{fig:data}) for the analysis was centered at (RA =307.17\degree, Dec=41.17\degree) with 6\degree{} radius and 2\degree{} mask around the brightest source in the region 2HWC J2019+367. This ROI  also includes 2HWC J2020+403, a gamma Cygni SNR which lies in the vicinity of 2HWC J2031+415. 

\begin{table}
\begin{center}
\caption{ Sources at the cocoon region}
\label{tab:sources}
\begin{tabular}{c c c c c} 

\textbf{Name}            & \textbf{Morphology}  &\textbf{RA }[deg]        &  \textbf{Dec} [deg]      &   \textbf{Source Association}\\      

2HWC J2020+403       & 0.63\degree{} radius disk          & 305.27           &        40.52          &        Gamma Cygni SNR   \\  

2HWC J2031+415    &    0.27\degree{} Gaussian width      & 307.89           &        41.58          &        VER J2031+415 PWN \\ 

HAWC J2030+406      &  2\degree{} Gaussian width        & 307.65           &        40.93          &        Fermi-LAT cocoon \\
\end{tabular}
\end{center}
\end{table}

\subsection{Sources at the Cocoon Region}
A multi-source model is developed to comprehensively describe the cocoon region and then fitted using the maximum likelihood code of 3ML \cite{3ml} . Both overlapping sources at 2HWC J2031+415 are described by 2D gaussian functions. One of the sources is a slightly extended region with the Gaussian width of 0.27\degree$\pm$0.03\degree{} co-located with VER J2031+415. The fit position (RA, Dec) is  (307.89\degree, 41.58\degree) as shown in Table \ref{tab:sources}. At HAWC energies, it is best described by a power law spectrum with an exponential cutoff as shown in Equation \ref{eq:plc}. The source is found to contribute $\sim$ 10\% flux to the total flux detected at the 2HWC J2031+415 region. 

\begin{equation}
\label{eq:plc}
\dfrac{dN}{dE}=N_{0}\left(\dfrac{E}{4.9{}\mathrm{TeV}}\right)^{\gamma} \cdot\ exp{\left(-E/E_{c}\right)},
\end{equation}
\
where $\gamma$ is spectral index and $E_{c}$ is cut off energy. \\
In VERITAS data, the emission of VER J2031+415, which is centered at the same location as 2HWC J2031+415 (within position uncertainties), is best described by asymmetric Gaussian morphology with width of -0.19\degree$\pm$0.02\degree by 0.08\degree$\pm$0.01\degree and a power law spectrum with spectral index of -2 \cite{veritas2014, ralph}. This source was first detected by the HEGRA observatory as an unidentified TeV emission and had been shown to favor a PWN model associated with PSR J2032+4127 by VERITAS \cite{aha02, veritas2014}.  To support the PWN hypothesis, VERITAS predicted that because of the hard index obtained for this source at VERITAS energies,  a cut-off around tens of  TeV must be observed due to the Klein-Nishina effect \cite{veritas2014}.  HAWC also detected a hard spectral index of -0.19$\pm$0.24 and obtained the cutoff $\left(E_{c}\right)$ at few tens of TeV favoring the PWN interpretation. However, HAWC detects a higher flux in comparison to the fluxes reported by VERITAS, HEGRA and MAGIC observatories. The morphology obtained for this source is $\sim$ 2--3 times larger with HAWC data, which could be one of the possible reasons for the detection of higher flux. Additionally, detailed studies are needed and ongoing to understand the multi-instruments morphology and spectrum for the source. \\

Both VER J2031+415 and 2HWC J2031+415 are situated well within an extended region of gamma-ray emission detected at GeV energies by Fermi-LAT. After subtracting the possible PWN source and the gamma Cygni SNR, as seen in Fig.~\ref{fig:cocoon} HAWC detects an extended region of gamma ray excess which is best fit by a Gaussian width of  2.18\degree$\pm$0.19\degree. The best fit position (RA, Dec) obtained is (307.65\degree $\pm$ 0.30\degree, 40.93\degree $\pm$ 0.26\degree).  The location and Gaussian width of the source agree with the measurements by Fermi-LAT (within statistical uncertainties). The energy spectrum of this TeV source is best descibed by a power law spectrum as shown in Equation \ref{eq:pl}. 
\begin{equation}
\label{eq:pl}
\dfrac{dN}{dE}=N_{0}\left(\dfrac{E}{4.2{}\mathrm{TeV}}\right)^{\gamma},
\end{equation}
\
where $\gamma$ is spectral index.\\
The large extended source contributed $\sim$  90\% to the total  flux detected at the 2HWC J2031+415 region. The spectral index measured by HAWC is -2.65$\pm$0.06, which is softer in comparison to the GeV spectrum. The same is true for ARGO J2031+4157 energy spectrum (spectral index of -2.6), which is located at (RA, DeC) = (307.8\degree $\pm$ 0.0.8\degree, 42.5\degree $\pm$ 0.6\degree) and has been suggested as a counterpart of the Cocoon at TeV energies \cite{Bartoli14}.  The flux measured by HAWC observatory agrees with the ARGO experiment.  The cocoon spectrum from GeV to TeV regime can be best described by a spectral shape that has a curvature.\\
 
\begin{figure}
\begin{subfigure}{.33\textwidth}
\captionsetup{width=.9\linewidth}
\includegraphics[scale=0.33]{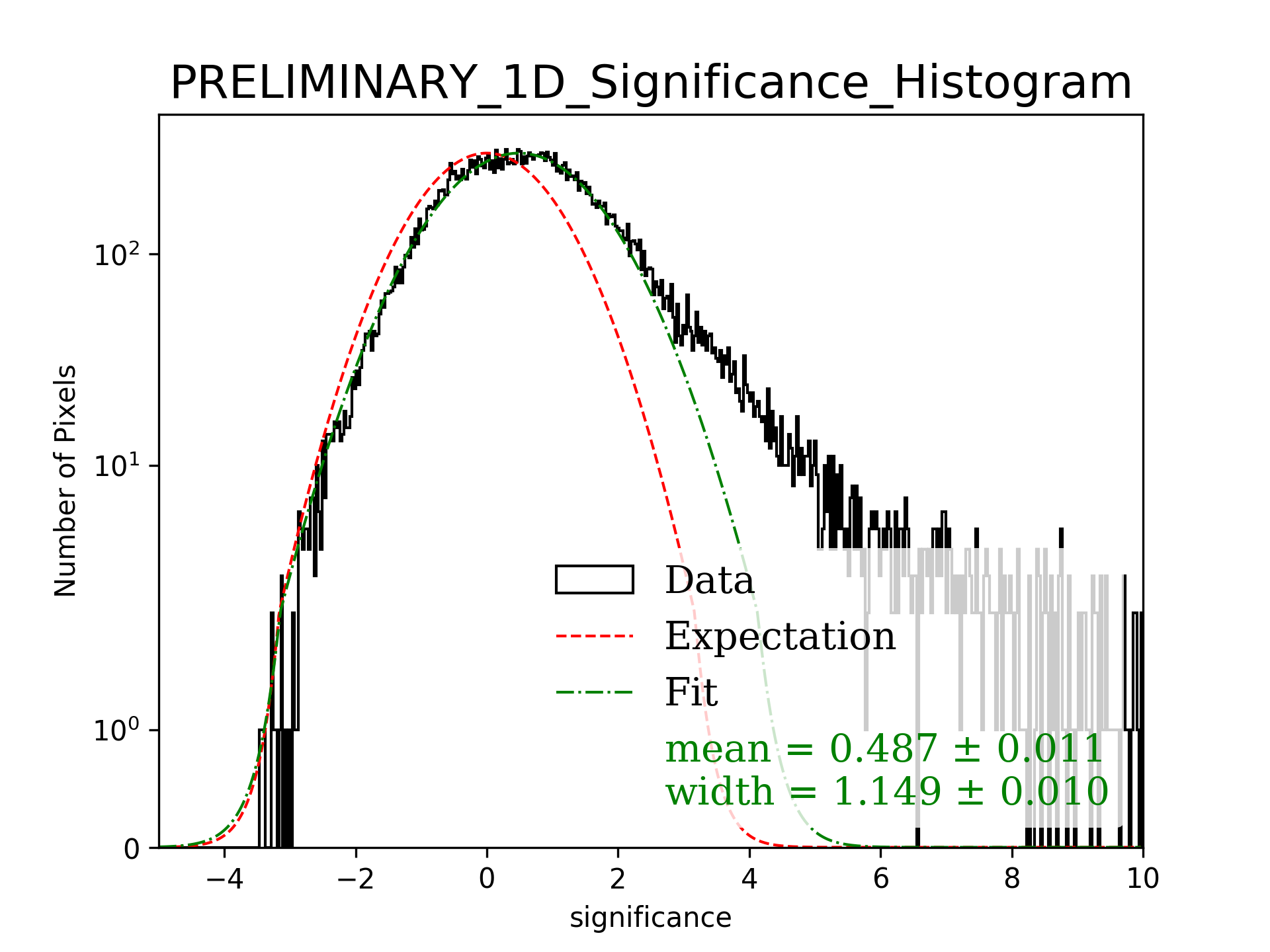}
\caption{Distribution of excesses before any source subtraction in the given ROI}
\label{fig:wholesig}
\end{subfigure}
\begin{subfigure}{.33\textwidth}
\captionsetup{width=.9\linewidth}
\includegraphics[scale=0.33]{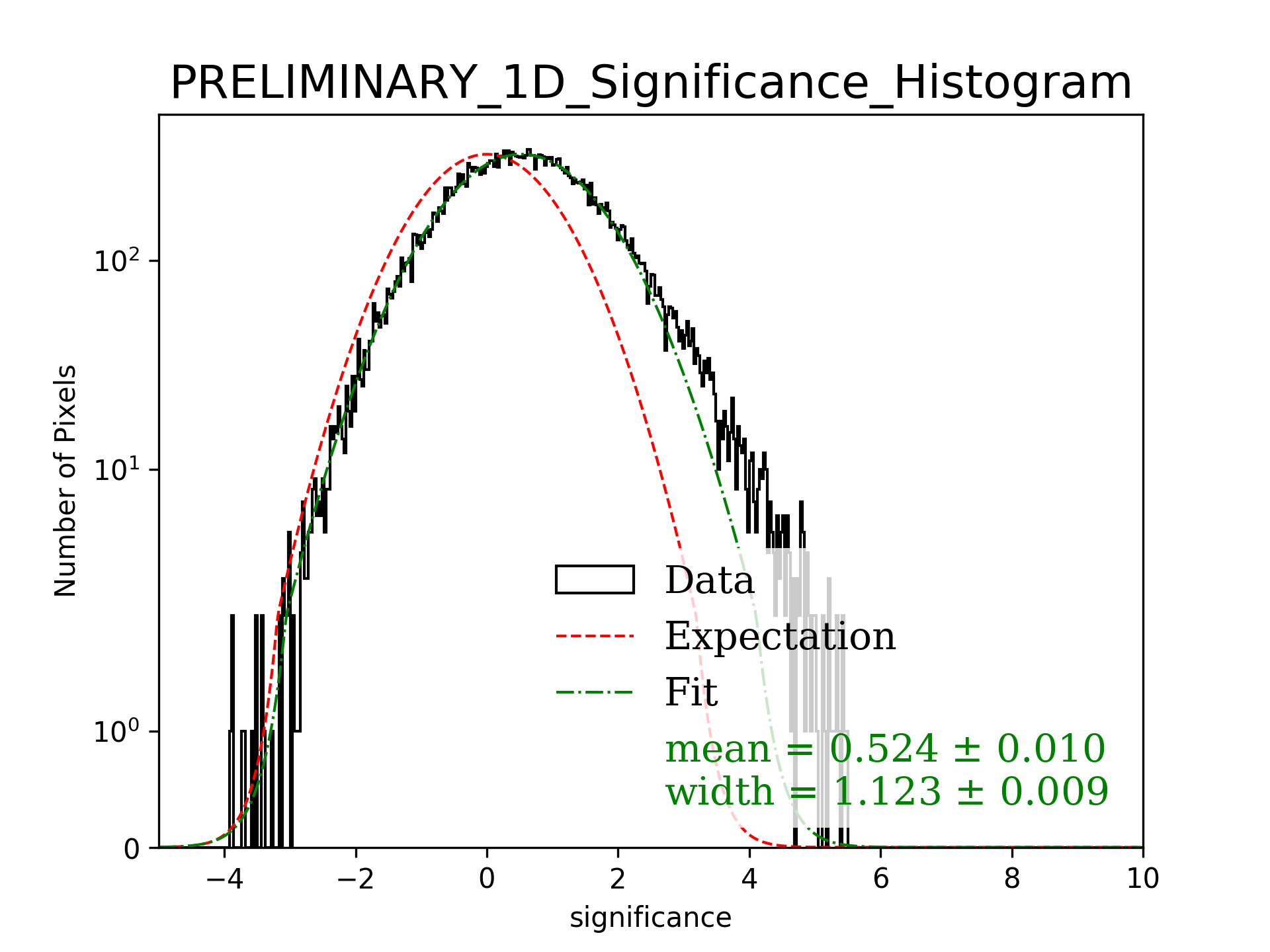}
\caption{Distribution of excesses after subtracting PWN source \& gamma Cygni}
\label{fig:sig1}
\end{subfigure}
\begin{subfigure}{.33\textwidth}
\captionsetup{width=.9\linewidth}
\includegraphics[scale=0.33]{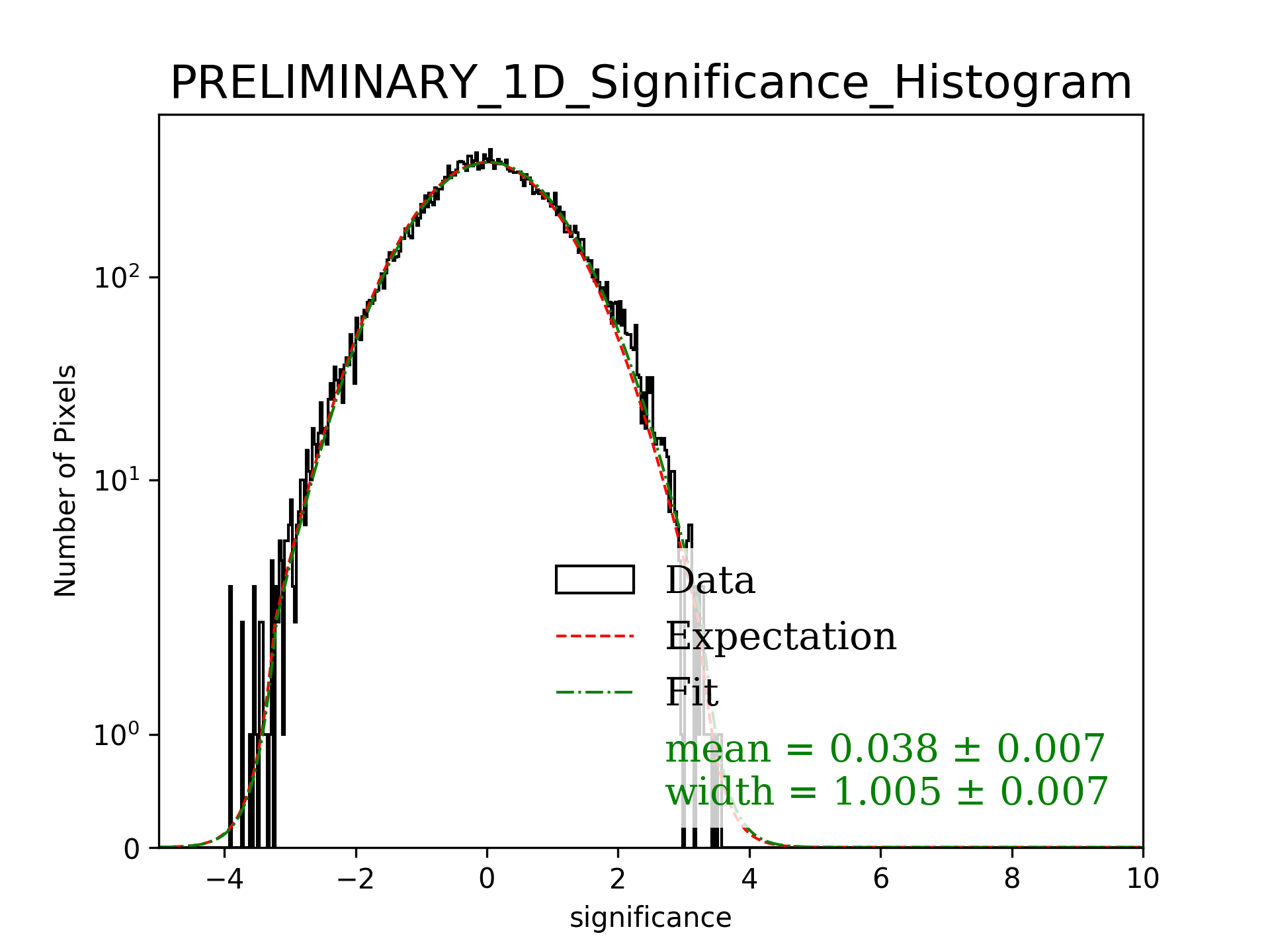}
\caption{Distribution of excesses after subtracting the best fit model in Table \ref{tab:sources} }
\label{fig:sig2}
\end{subfigure}
\captionsetup{width=.9\linewidth}
\caption{Significance distribution in the analysis ROI}
\end{figure}

The best fit model of the region in the ROI  also includes the gamma Cygni SNR, which is described by an extended disk of  $\sim$ 0.63\degree{} and a power law spectrum and is described in more detail in \cite{henrike}.  To check for a diffuse emission component in the region, it was included as a uniform or a Gaussian background in addition to three gamma-ray sources . In both cases, the diffuse emission was not detected significantly and adding the component decreased cocoon flux by $\sim$ 10 - 15 \%.  This is a contrast to the Fermi-LAT energy range, where diffuse emission is important and a dedicated template was developed for the region.  Fig.~\ref{fig:wholesig} shows the distribution of excess significance of the data in the analysis ROI. Fig.~\ref{fig:sig1} shows the distribution of excesses after subtracting PWN and gamma Cygni. Since the emission from the cocoon source has not been subtracted, the distribution is skewed towards the positive values. As can be seen in the Fig.~\ref{fig:sig2}, after subtraction of all three sources, there is no longer excess above background fluctuations. This supports the hypothesis that the current best fit model is a good description of the region and accounts for the total TeV gamma-ray emission observed at the cocoon region by HAWC.

\section{Conclusion and Discussion}
SFRs can accelerate CRs through the collective effects of stellar winds and additional SN explosion of old massive stars. A few of these regions have been observed in gamma rays. One such region is the Fermi-LAT cocoon, which was first detected at GeV energies, now has a possible TeV counterpart. The study of the cocoon region with HAWC data shows that after subtraction of the other sources in the region, the likely TeV counterpart is significantly detected. This TeV emission observed by HAWC is  morphologically similar in nature to its GeV counterpart. In both GeV and TeV regime, the spectrum is best described by a power law function. However, while at GeV energies, the spectrum is harder with $\sim$ -2.1 index, the spectrum softens at TeV energies with an index of $\sim$ -2.6 and provides an evidence of spectral curvature from GeV--TeV regime.\\

The cocoon lies in the low gas density superbubble. The OB2 association can accelerate CRs which escape and interact with the gas cloud. The confinement of the CRs in the superbubble structure supports the proposed mechanism of acceleration by the collective stellar winds of the massive stars in the OB2 association. The large extended region of both GeV and TeV emission and the lack of X-ray counterparts support the possibility of hadronic origin. 

Assuming the hadronic scenario, with HAWC data, there is evidence of CR acceleration to hundreds of TeV, with a lower limit of a few hundred TeV for the maximum CR energy.  For a hadronic scenario, assuming the gas density of 30 nucleons/${cm^3}$ \cite{butt}, the GeV--TeV spectral fit with HAWC data and Fermi-LAT flux points results in the total CR energy above 1 GeV in the cocoon region to be $\sim$ $10^{50}$ erg. In the OB2 region, wind power of $\sim$ $10^{39}$ erg/sec has been maintained continuously over the past $\sim$ 2Myrs \cite{OB2energy}, giving the total energy in the region due the stellar winds $\sim$ $6 \cdot 10^{52}$ erg. Calculating the fraction of the total CR energy of the cocoon to the total energy budget of wind power in the OB2 association results in an acceleration efficiency of $\sim$ 0.1\%. Thus, the association has a high energy budget to account for the acceleration of CRs that now make up the cocoon.\\

In addition to the preliminary hardronic scenario mentioned here, both leptonic and hadrodic models are being investigated. While, in the preliminary studies, there is support for hadronic scenario, a combination of both leptonic and hadronic emission can not be ruled out yet. With these further  detailed studies, we plan to confirm the origin nature of the emission and better constrain the maximum CR energy in the emission region.

\end{document}